\documentclass[12pt]{article}

\usepackage{amssymb}
\usepackage{amsmath}
\usepackage{amscd}
\usepackage{latexsym}
\usepackage{cite}

\usepackage{graphicx}

\usepackage{subfigure}

\newcommand{\be}{\begin{equation}}
\newcommand{\ee}{\end{equation}}

\newcommand{\prt}{\partial}
\newcommand{\br}{{\bf r}}

\newcommand{\vp}{\varphi}

\newcommand{\gm}{\gamma}
\newcommand{\om}{\omega}

\begin{document}

\begin{center}

{\Large{\bf Formation of granular structures \\ 
in trapped Bose-Einstein condensates \\
under oscillatory excitations}} \\ [5mm]

V.I. Yukalov$^{1,2*}$, A.N. Novikov$^{1}$, and V.S. Bagnato$^2$ \\ [3mm]

{\it
$^1$Bogolubov Laboratory of Theoretical Physics, \\
Joint Institute for Nuclear Research, Dubna 141980, Russia \\ [3mm]

$^2$Instituto de Fisica de S\~{a}o Calros, Universidade de S\~{a}o Paulo, \\
CP 369, 13560-970 S\~{a}o Carlos, S\~{a}o Paulo, Brazil} \\ [3mm]
\end{center}

\vskip 3cm

\begin{abstract}
We present experimental observations and numerical simulations of nonequilibrium
spatial structures in a trapped Bose-Einstein condensate subject to oscillatory 
perturbations. In experiment, first, there appear collective excitations, followed 
by quantum vortices. Increasing the amount of the injected energy leads to the 
formation of vortex tangles representing quantum turbulence. We study what happens 
after the regime of quantum turbulence, with increasing further the amount of 
injected energy. In such a strongly nonequilibrium Bose-condensed system of trapped 
atoms, vortices become destroyed and there develops a new kind of spatial structure 
exhibiting essentially heterogeneous spatial density. The structure reminds fog 
consisting of high-density droplets, or grains, surrounded by the regions of low 
density. The grains are randomly distributed in space, where they move. They live 
sufficiently long time to be treated as a type of metastable objects. Such structures 
have been observed in nonequilibrium trapped Bose gases of $^{87}$Rb, subject to 
the action of alternating fields. Here we present experimental results and support 
them by numerical simulations. The granular, or fog structure is essentially 
different from the state of wave turbulence that develops after increasing further
the amount of injected energy.   
\end{abstract}

\vskip 0.5cm

$^*$Corresponding author (V.I. Yukalov):

{\bf E-mail}: yukalov@theor.jinr.ru

\newpage

\section{Granular structure characteristics}

A granular structure is understood as an inhomogeneous mixture of sufficiently large 
dense formations, called grains, surrounded by a much less dense phase, such as gas. 
That is, a granular structure does not constitute a single phase of matter, but has 
the properties of two or more intermixed phases. Such granular materials are ubiquitous 
in nature, usually being a mixture of solid formations separated by gaseous phase 
\cite{Bagnold_1,Duran_2,Pudasaini_3}.

In the present paper, we consider a mixture enjoying similar properties, though being 
very different from the usual granular materials. It is also an inhomogeneous 
composition of two phases, more dense and essentially less dense one. Although the 
more dense phase is not solid. The main difference of the mixture we shall consider 
is that both phases, the dense and rarified one, are formed by the same type of atoms. 
Such a mixture can be created in a cloud of trapped Bose-condensed atoms and in 
optical lattices. Before describing the concrete case of trapped atoms, let us 
emphasize the general features of the granular mixture that we will be considering 
in what follows. There are five main properties characterizing such a granular 
structure.

(i) The structure is composed of the same type of atoms that form two or more different 
phases. As a whole, the system is not equilibrium, but it has to be in a locally 
equilibrium state in order that it could be possible to speak of different phases 
forming the mixture. This requirement distinguishes the mixture, we shall consider, 
from the usual granular materials, such as sand, coal, rice, coffee, or corn flakes, 
whose grains are formed of different atoms as compared to the surrounding air. To be 
in local equilibrium the typical grain lifetime $t_g$ has to be much larger than the 
local equilibration time $t_{loc}$.

(ii) The spatial distribution of grains at a snapshot is random. They constitute no 
ordered structure, such as domains, stripes or other patterns.

(iii) The spatial locations of grains in different experiments is also random, so 
that there is no repeating spatial structure, but they are randomly distributed. 

(iv) The typical linear size of grains, $l_g$, is mesoscopic, being between the 
scattering length $a_s$ of atoms composing a grain and the total system size $L$, 
so that $a_s \ll l_g \ll L$.

(v) The grains are of multiscale nature, having the sizes in a dense interval 
$[l_{min}, l_{max}]$ and possessing different shapes. This means that the mentioned 
typical size $l_g$ is not just a single fixed quantity, but an average typical 
quantity from an interval $[l_{min}, l_{max}]$. 

These features make it rather difficult to describe such a granular mixture composed 
of grains that are random in shapes, sizes, and in spatial distribution. In addition, 
to create a granular mixture of the considered type it is, generally, necessary to 
strongly perturb the atomic system, moving it far from equilibrium.

\section{Perturbing trapped condensate}

In this section, we delineate the general sequence of states through which the system 
of trapped condensed atoms passes in the process of their excitation by an external 
field, when gradually increasing its strength and time of action. A concrete experiment 
with $^{87}$Rb will be described in Sec. 3 and numerical simulations discussed in Se. 4. 

Trapped Bose atoms in equilibrium at low temperatures form Bose-Einstein condensate 
in the ground state. The condensate cloud in a trap enjoys approximately Thomas-Fermi 
shape, with well known properties described in the books 
\cite{Pitaevskii_4,Lieb_5,Letokhov_6,Pethick_7} and reviews \cite{Courteille_8,Andersen_9,Yukalov_10,Bongs_11,Yukalov_12,Posazhennikova_13,Yukalov_14,
Proukakis_15,Yurovsky_16,Yukalov_17,Yukalov_18}. In order to strongly perturb the 
condensate, it is necessary to impose external perturbations transferring the condensate 
from its ground state to excited states. There are two main ways of imposing such 
external perturbations.

One possibility is to add to the static trapping potential $U({\bf r})$ an alternating 
potential $V({\bf r},t)$, so that the total trap potential becomes
\be
\label{1}
 U(\br,t) = U(\br) + V(\br,t) \;  .
\ee
Another way is to modulate the scattering length $a_s(t)$ by means of Feshbach 
resonance techniques. Both these ways can be used for strongly disturbing Bose-Einstein condensates.

Suppose trapped Bose atoms have been cooled down to very low temperatures, when 
practically all of them pile down to a Bose-condensed state. And let us apply an 
external modulating perturbation by one of the methods mentioned above. First, at 
weak perturbation, there appear elementary collective excitations that are small 
deviations from the ground state. Weak perturbations also can generate large deviations 
from the ground state, provided that the modulation frequency is in resonance with 
one of the transition frequencies between topological coherent modes 
\cite{Yukalov_19,Yukalov_47}. The latter are defined as the eigenfunctions of the 
nonlinear stationary Schr\"{o}dinger equation
\be
\label{2}
 \left [ -\; \frac{\nabla^2}{2m} + U(\br) + N \Phi_0 | \vp_n(\br)|^2
\right ] \vp_n(\br) = E_n \vp_n(\br) \;  ,
\ee
where $N$ is the number of condensed atoms, assumed to be close to the total number 
of atoms, and
\be
\label{3}
\Phi_0 \equiv 4\pi\; \frac{a_s}{m}
\ee
is the atomic interaction strength, in which $a_s$ is a scattering length assumed 
to be positive. The modes are termed topological, since they have different number 
of zeroes, thus, topologically different atomic density. They are coherent, being 
formed by condensed atoms characterized by a coherent state.

In the above equations, the Planck constant is set to unity. While we shall restore 
it below for the clarity of numerical estimates.

The known particular example of the topological modes are quantum vortices. If the 
external perturbation rotates the atomic cloud, acting as a spoon, then vortices 
appear being aligned along the imposed axis of rotation. But when the trap modulation 
does not prescribe a fixed rotation axis, then vortices and antivortices arise in pairs 
or in larger groups \cite{Courteille_8,Yukalov_19}. The explicit experimental 
demonstration for the appearance of clusters of vortices and antivortices was done 
in Ref. \cite{Seman_20}.   

Increasing the strength of the trap modulation generates a variety of coherent modes, 
needing no resonance conditions because of the power broadening effect 
\cite{Yukalov_21}. Among these numerous coherent modes, the basic vortex, with the 
winding number one, is the most energetically stable. For a trap with a transverse, 
$\omega_{\perp}$, and longitudinal, $\omega_z$, frequencies, the vortex energy can 
be written \cite{Pethick_7} as
\be
\label{4}
 \om_{vor} = \frac{0.9\om_\perp}{(\nu g)^{2/5} } \;
\ln ( 0.8 \nu g ) \;  ,
\ee
where the notation is used for the trap aspect ratio
\be
\label{5}
\nu \equiv \frac{\om_z}{\om_\perp} = \left ( \frac{l_\perp}{l_z}
\right )^2
\ee
and for the effective coupling parameter
\be
\label{6}
 g \equiv 4\pi N \; \frac{a_s}{l_\perp} \; ,
\ee
with $l_{\perp}$ and $l_z$ being the transverse and longitudinal oscillator lengths,
respectively. Due to the large number of atoms $N$, the effective coupling parameter 
is large, $g \gg 1$. As is seen, the basic vortex energy diminishes with the increase 
of $g$. At the same time, the transition frequencies of other modes, hence their 
energies, can be shown \cite{Yukalov_19,Yukalov_21} to increase as
\be
\label{7}
 \om_{mn} \; \propto \; (\nu g)^{2/5} \qquad ( g \gg 1) \; .
\ee
This makes the basic vortex the most energetically stable mode.

When the trap aspect ratio is not too small, the trap can house many vortices. The 
latter are created due to dynamic instability arising in the moving fluid 
\cite{Yukalov_22,Dutton_23,Yukalov_24,Ruostekoski_25,Shomroni_26,Ma_27,Ishiro_28,Simula_29}.

Increasing the strength of the pumping, without imposing any rotation axis, produces 
a tangle of vortices, which makes the trapped atomic cloud turbulent 
\cite{Henn_30,Seman_31,Shiozaki_32,Seman_33,Bagnato_48}. Increasing further either 
the amplitude of the pumping field or the pumping time leads to the appearance of 
different structures, such as the granulated condensate state.

The energy, injected into the trap by means of an external perturbation modulating 
the trapping potential, can be written as
\be
\label{8}
 E_{inj} = \int \rho(\br,t) \left |\frac{\prt V(\br,t)}{\prt t} \right | \; d\br dt \;  ,
\ee
where $\rho({\bf r}, t)$ is atomic density. In the case of a periodic in time 
alternating field $V({\bf r},t) \sim A \cos(\omega t)$, the energy, injected during 
the modulation time $t$, takes the form $E_{inj} \approx \omega t$. This makes it 
possible to represent the crossover lines between different regimes as the relation
\be
\label{9}
A \sim \frac{\pi E_{inj}}{2\om t}
\ee
between the amplitude $A$ of the pumping field and the pumping time $t$.

The system properties also depend on pair atomic interactions that are conveniently 
characterized by the gas parameter
\be
\label{10}
 \gm \equiv \rho^{1/3} a_s = \frac{a_s}{a} \;  ,
\ee
which is usually small for trapped atoms, although can be varied in a wide range 
by means of the Feshbach resonance techniques. Despite the gas parameter $\gamma$ 
can be small, but the effective coupling parameter $g$ is usually large because of 
the large number of atoms in a trap. 

An important quantity, showing whether atoms are in local equilibrium, is the local 
equilibration time
$$
 t_{loc} \equiv \frac{m}{\hbar\rho a_s} = 4\pi \;
\frac{m\xi^2}{\hbar} \;  ,
$$
where $\xi \equiv 1/\sqrt{4 \pi \rho a_s}$ is the coherence length. A perturbed cloud 
of trapped atoms, as a whole, can be strongly nonequilibrium, while, at the same time, 
be locally equilibrium. This happens in the situation, when the modulation period 
$t_{mod} \equiv 2 \pi /\omega$ of the alternating modulating field, with frequency 
$\omega$, is much longer than the local equilibration time $t_{loc}$.

\section{Modulation of trapping potential for $^{87}$Rb}

In experiments, the granular condensate structures are created for trapped 
$^{87}$Rb by modulating the trap with a time alternating potential.

The cloud of $^{87}$Rb atoms, of mass $m = 1.443 \times 10^{-22}$ g and scattering 
length $a_s = 0.557 \times 10^{-6}$ cm, is cooled down to temperatures much lower 
than the Bose-Einstein condensation temperature $T_c = 276$ nK, so that the great 
majority of all $N = 2 \times 10^5$ atoms are condensed, with the condensate 
fraction being $n_0 = 0.7$.

The trap is of cylindrical shape, with the frequencies 
$\om_\perp = 2 \pi \times 210$ Hz and $\om_z = 2 \pi \times 23$ Hz, which corresponds 
to the oscillator lengths $l_\perp = 0.74 \times 10^{-4}$ {\rm cm} and 
$l_z = 2.25 \times 10^{-4}$ {\rm cm}. Respectively, the trap aspect ratio is 
$\nu = 0.11$. The effective coupling parameter is $g = 1.96 \times 10^4$.

The atomic cloud is characterized by the sizes that can be estimated as  
$r_\perp = 2.27 \times 10^{-4} {\rm cm}, z_0 = 1.47 \times 10^{-3} {\rm cm}$, which 
makes it possible to find the effective atomic density 
$\rho = 0.43 \times 10^{15}$ cm$^{-3}$ and the mean interatomic distance 
$a = 1.32 \times 10^{-5}$ cm. The gas parameter is small, 
$\gamma \sim 1.4 \times 10^{-3}$, while the effective coupling is large, 
$g = 1.96 \times 10^4$. 

The trap potential is modulated by an additional alternating potential 
$V({\bf r}, t)$ (see \cite{Seman_20,Henn_34,Bagnato_48}) oscillating with 
the frequency $\omega = 1.26 \times 10^3$ s$^{-1}$, which corresponds to the 
modulation period $t_{mod} = 0.5 \times 10^{-2}$ s. The total modulation time 
$t_{ext}$ is varied between $0.02$ s and $0.1$ s. The local equilibration time is 
$t_{loc} = 0.57 \times 10^{-3}$ s. Thus, the relations between the characteristic 
times is
$$
 t_{loc} \ll t_{mod} \ll t_{ext} \;  ,
$$
which means that the system is locally equilibrium, although as a whole it is 
strongly nonequilibrium.

Because of the high atomic density inside the trap, the {\it in situ} observation 
is impossible. Absorption pictures are taken in the time-of-flight setup, after the 
times $t_{tof}$ between $0.015$ s and $0.023$ s. The turbulent and granular structures, 
consequently created by increasing the excitation time, are displayed in Fig. 1. 
The final stage, corresponding to the granular state, is shown in details in Fig. 2. 
Since the image is obtained through the absorption by a three-dimensional expanding 
cloud, the contrast between the grains and rarified gas surrounding them is not very 
sharp. Restoring the characteristic linear size of grains before the free expansion, 
we get $l_g \approx 3 \times 10^{-5}$ cm. The relation between the characteristic 
lengths is 
$$
 a_s \ll a \sim \xi \sim l_g \ll l_z \; .
$$
This shows that the grains are of mesoscopic size, being in between the scattering 
length and the linear system size. At the same time, the grain linear size is close 
or slightly larger than the coherence length, which suggests that each grain 
represents a coherent formation.   

The excitation of strongly nonuniform states can also be realized by modulating 
the scattering length \cite{Ramos_35,Pollack_36}. Generally, long modulation times 
or large exciting amplitudes generate the cloud evolution from the appearance of 
separate vortices to tangled vortex configurations, typical of quantum turbulence,
after which the granular state arises. 

The experimental phase diagram on the amplitude-time $A-t$ plane is described in 
Refs. \cite{Shiozaki_32,Seman_33}, where it is shown that with increasing the 
injected energy, that is proportional to the product $A t$, the system passes 
through the following states: {\it regular superfluid} slightly perturbed by a 
random weak external field, {\it vortex superfluid} with several vortices, 
{\it turbulent superfluid} formed by a tangle of many vortices, and 
{\it granular state} with condensate droplets surrounded by a gas of low density.

\section{Numerical simulations for $^{87}$Rb experiment}

In order to better understand the properties of the granular sate, we have 
accomplished numerical simulations for exactly the same setup and parameters as
in experiment \cite{Shiozaki_32,Seman_33,Bagnato_48} with $^{87}$Rb.

The simulation is based on the nonlinear Schr\"{o}dinger equation, with the 
parameters corresponding to the experiments \cite{Shiozaki_32,Seman_33,Bagnato_48}. 
The same alternating potential modulating the trap as in these experiments is used. 
The amount of the energy pumped into the trap, as is illustrated in Eq. (8), is 
proportional to the amplitude of the modulating potential and to the pumping time, 
approximately through the product of these. Increasing $E_{inj}$, we in turn observe 
first, a slightly nonequilibrium regular Bose-condensed atomic cloud, then the 
appearance of separate vortices. After the number of vortices reaches about 25, the 
regime of quantum vortex turbulence develops corresponding to the random vortex tangle. 
These regimes of the {\it regular superfluid}, {\it vortex state}, and 
{\it vortex turbulence} have been thoroughly described in our previous papers 
\cite{Shiozaki_32,Seman_33,Bagnato_48}. Therefore here we pay the main attention to 
the granular state that arises after the regime of vortex turbulence. 

The increase of the injected energy $E_{inj}$ leads, after the vortex turbulence, to 
the appearance of the granular state, when there are no vortices, but the system 
decomposes into a number of dense grains, or droplets, inside a rarified surrounding. 
The density of the grains is up to 100 times larger than that of their surrounding. 
The order-parameter phase is practically the same inside each grain, confirming that 
these grains are the droplets of Bose-condensed atoms. But the phases inside different 
grains can be different. The grains differ from vortices by the absence of their
vorticity. A typical granular structure is illustrated in Fig. 3.

The sizes of the grains vary in the range between $1 \times 10^{-5}$ cm to 
$3 \times 10^{-5}$ cm. Although they are not spherical, but their linear sizes in the 
radial and longitudinal directions are close to each other. 

After the granular state has been created, we have analyzed its stability by switching 
off the perturbing potential and observing the spatio-temporal behavior of the system. 
The grains can move in space, slightly changing their shapes, sometimes fusing with 
each other, but survive during the period of time of order $10^{-2}$ s. The grain 
lifetime for $^{87}$Rb is in agreement with the estimate $t_g \sim (\xi / a_s) t_{loc}$ 
for the heterophase lifetime \cite{Yukalov_40} giving for $^{87}$Rb the same order 
of $10^{-2}$ s. The grain lifetime is much longer than the local-equilibration time 
$t_{loc} \sim 10^{-4}$ s, which proves that the grains can be treated as 
quasi-equilibrium formations. The spatio-temporal behavior of the grains is 
demonstrated in the sequence of the transverse cross-sections in Fig. 4. 

In each realization, either numerical or experimental, the spatial distribution of 
grains is random and does not repeat from one realization to another. The regime 
corresponding to this random granular state can be termed {\it grain turbulence}. Since 
the system in this state consists of two types of regions of drastically different 
density, hundreds of time differing from each other, such a state is an example of 
a {\it heterogeneous}, or {\it heterophase}, state. Therefore the grain turbulence 
is a particular case of heterophase turbulence \cite{Yukalov_41}.

Finally, increasing the amount of injected energy, after the regime of the grain 
turbulence, another state develops, consisting of uniformly distributed in space weak 
waves, whose density is only about 3 times larger than that of their surrounding. The 
typical linear sizes of separate waves are between $0.3 \times 10^{-4}$ cm to 
$0.8 \times 10^{-4}$ cm. The phases inside the waves as well as between them are random. 

This wave regime corresponds to the so-called {\it wave turbulence}, or 
{\it weak turbulence}. The latter is principally different from the regime of 
{\it grain turbulence}. The regime of weak turbulence is demonstrated in Fig. 5 as 
the sequence of density snapshots. Our simulations show that the regimes of grain 
turbulence and wave turbulence are clearly distinguished by the following main features:

\vskip 2mm

(i) The waves are weak, having the amplitudes only about 3 times larger than the most 
rarified parts of the system, while the grains are dense formations whose density is 
up to 100 times larger than that of their surrounding.

\vskip 2mm

(ii) The phase inside each wave is rather random, while the phase inside a grain 
is practically constant. This implies that the wave turbulence, actually, corresponds 
to the situation when the condensate is destroyed in the whole system, while grain 
turbulence describes the intermediate case, where there exist coherent Bose condensate 
germs, or droplets, inside a rarified normal phase.     

\vskip 2mm

(iii) In the regime of wave turbulence, the system kinetic energy is more than 100 
times larger than the interaction energy, while under grain turbulence the former is 
only about five times larger than the latter. 
        
\vskip 2mm

In this way, wave turbulence, we reproduce in numerical simulations, completely 
satisfies the commonly accepted basic features of this regime \cite{Zakharov_42}. 
And the grain turbulence corresponds to the intermediate regime of random turbulent 
cells in the Kibble-Zurek \cite{Kibble_43,Zurek_44} picture of transition between 
the vortex turbulence and normal turbulent state. It is possible to show 
\cite{Yukalov_45} that there exists a mapping between the states of an atomic bosonic 
cloud, subject to an alternating trap modulation, and the states of an atomic system 
in a random spatial external potential \cite{Yukalov_46}. According to this mapping, 
grain turbulence corresponds to a disordered Bose system and the wave turbulence, 
to the normal state with destroyed Bose condensate.

\section{Conclusion}

We have presented experimental evidence for the formation of granulated Bose-Einstein 
condensate, consisting of dense condensate droplets immersed into the surrounding 
rarified gas. The density of the grains can be hundred times larger than that of the 
surrounding gas. Such granulated Bose condensates have been observed in nonequilibrium 
trapped Bose gases of $^{87}$Rb, under alternating trap modulation. 

Numerical simulations, based on the nonlinear Schr\"{o}dinger equation, with the 
parameters exactly corresponding to the experiments with $^{87}$Rb, reproduce well all 
the stages of the condensate excitation by means of the trapping potential alternation.
Increasing the amount of energy, injected into the system by the trap modulation, we 
observe in turn the following states: slightly perturbed 
{\it regular Bose-Einstein condensate}, {\it vortex state}, with a few vortices and 
antivortices, {\it vortex turbulence} formed by a random tangle of quantum vortices, 
{\it grain turbulence} represented by randomly distributed grains of dense Bose-condensate
droplets inside a very rarified gas, and {\it wave turbulence} consisting of the 
random waves of small amplitude, where spatial coherence is destroyed.   

Our numerical simulations confirm that the granular state can be considered as a 
metastable state, sice the lifetime of the coherent grains is much longer that the 
local equilibration time. The existence and properties of granulated structures 
essentially depend on the system parameters.

In a recent experiment performed in the laboratory of R. Hullet, using $^{7}$Li and 
modulation of interaction, it seems there has also been observed the formation of 
granular structure. This is currently under investigation and will be the topic of 
a future publication.

\vskip 5mm

{\bf Acknowledgement}

\vskip 3mm

Financial support from FAPESP and CNPQ (Brazil) is appreciated. Two of the authors 
(A.N.N. and V.I.Y.) acknowledge financial support from the Russian Foundation for 
Basic Research. One of the authors (V.I.Y.) is grateful for discussions to E.P. Yukalova.

\newpage

\begin{center}

{\Large{\bf Figure Captions } }

\end{center}

\vskip 1cm

{\bf Fig. 1}. Image of the full sample observed by absorption in experiments 
with $^{87}$Rb. The turbulent cloud of vortices (top) demonstrates the random vortex 
tangle, while the granular state (bottom) consists of the domains of very different 
atomic density, reminding droplets. The granulation appears when perturbing the system
for long time, after passing through the stage of vortex turbulence.

\vskip 1cm

{\bf Fig. 2}. Details of the observed absorption image of a granulated atomic 
superfluid of $^{87}$Rb. The domains of high density of a variety of sizes and shapes 
are seen.

\vskip 1cm

{\bf Fig. 3}. The granular structure of $^{87}$Rb realized in numerical modeling. 
The grains (droplets) are clearly seen in the density snapshots.

\vskip 1cm

{\bf Fig. 4}. Spatio-temporal behavior of the grains illustrated by numerical 
simulations. Each column represents the sequence of transverse cross-sections of the 
atomic cloud at different relaxation times $\tau = 0; 1,5; 3$ and 5 ms 
(from left to right), after the perturbing potential is switched off. The granular 
structure becames blurred during the time, but still well observable after 5 ms.

\vskip 1cm

{\bf Fig. 5}. Transverse cross-sections of the $^{87}$Rb atomic cloud, 
corresponding to the regime of wave turbulence, as found in numerical simulations. 

\newpage

\begin{figure}[ht]
\centering
\includegraphics[width=10cm]{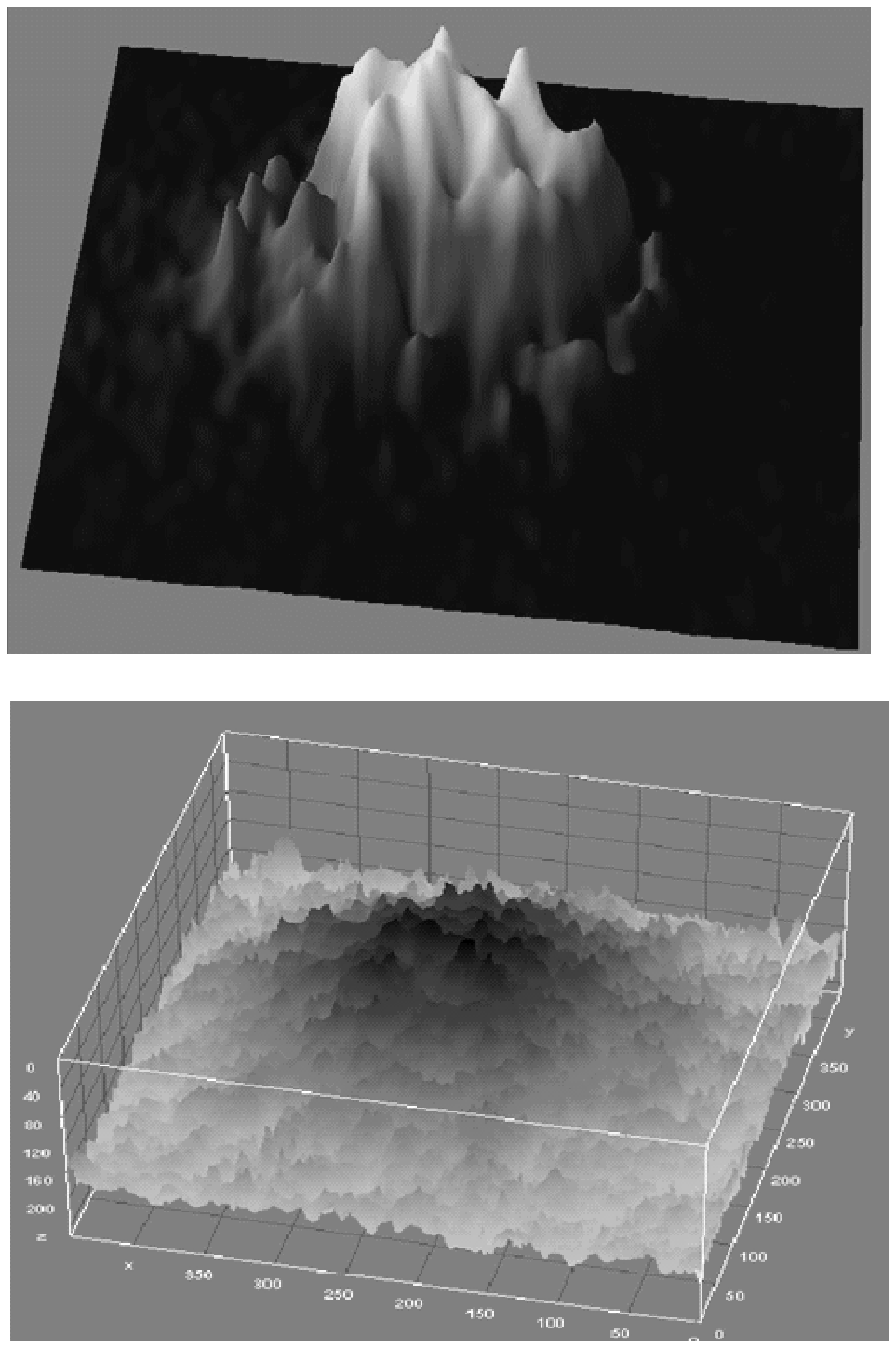}
\caption{Image of the full sample observed by absorption in experiments 
with $^{87}$Rb. The turbulent cloud of vortices (top) demonstrates the random vortex 
tangle, while the granular state (bottom) consists of the domains of very different 
atomic density, reminding droplets. The granulation appears when perturbing the system
for long time, after passing through the stage of vortex turbulence.}
\end{figure}

\vskip 2cm

\begin{figure}[ht]
\centering
\includegraphics[width=10cm]{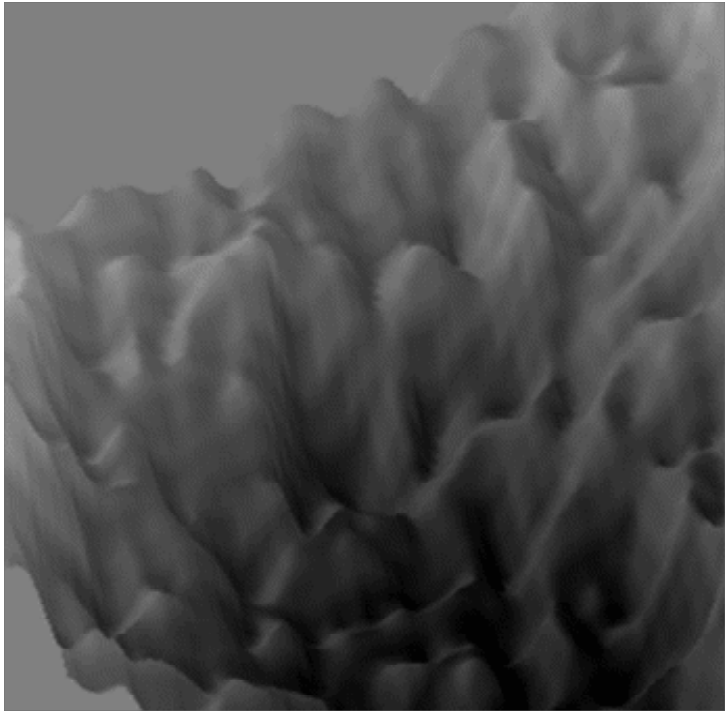}
\caption{Details of the observed absorption image of a granulated atomic 
superfluid of $^{87}$Rb. The domains of high density of a variety of sizes and shapes 
are seen.}
\end{figure}

\vskip 2cm

\begin{figure}[!ht]
\centering
\includegraphics[width=10cm]{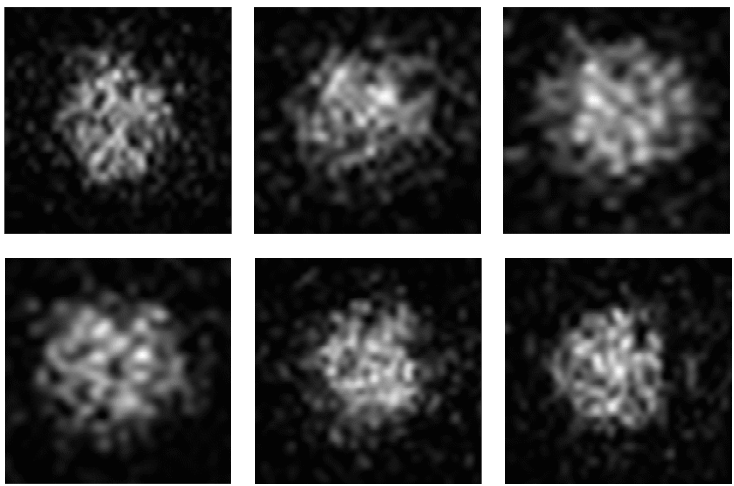}
\caption{The granular structure of $^{87}$Rb realized in numerical modeling. 
The grains (droplets) are clearly seen in the density snapshots.}
\end{figure}

\newpage

\begin{figure}[!ht]
\centering
\includegraphics[width=10cm]{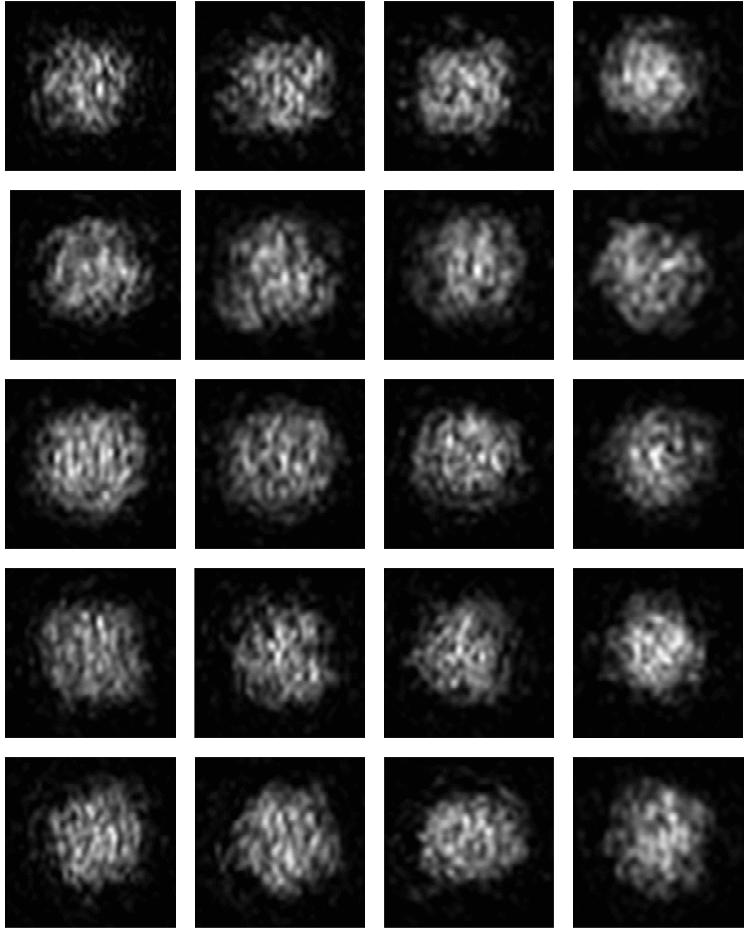}
\caption{Spatio-temporal behavior of the grains illustrated by numerical 
simulations. Each column represents the sequence of transverse cross-sections of the 
atomic cloud at different relaxation times $\tau = 0; 1,5; 3$ and 5 ms 
(from left to right), after the perturbing potential is switched off. The granular 
structure becomes blurred during the time, but still well observable after 5 ms.}
\end{figure}

\begin{figure}[!ht]
\centering
\includegraphics[width=10cm]{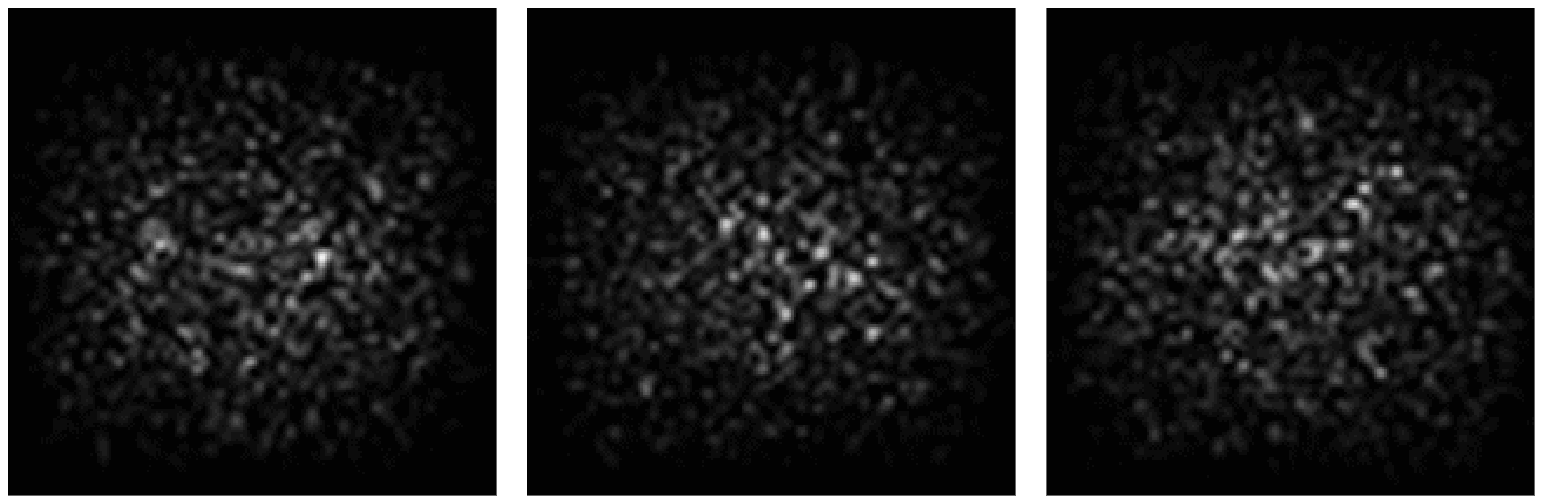}
\caption{Transverse cross-sections of the $^{87}$Rb atomic cloud, 
corresponding to the regime of wave turbulence, as found in numerical simulations.} 
\end{figure}


\newpage

\begin{thebibliography}{99}


\bibitem{Bagnold_1}
Bagnold R A 1941 
{\it The Physics of Blown Sand and Desert Dunes} 
(London: Methuen)

\bibitem{Duran_2}
Duran J 1999 
{\it Sands, Powders, and Grains: An Introduction to the Physics of Granular Materials} 
(New York: Springer)

\bibitem{Pudasaini_3}
Pudasaini S P and Hutter K 2007 
{\it Avalanche Dynamics: Dynamics of Rapid Flows of Dense Granular Avalanches} 
(Berlin: Springer)

\bibitem{Pitaevskii_4}
Pitaevskii L and Stringari S 2003 
{\it Bose-Einstein Condensation} 
(Oxford: Clarendon)

\bibitem{Lieb_5}
Lieb E H, Seiringer R, Solovej J P and Yngvason J 2005 
{\it The Mathematics of the Bose Gas and Its Condensation} 
(Basel: Birkhauser)

\bibitem{Letokhov_6} 
Letokhov V 2007 
{\it Laser Control of Atoms and Molecules} 
(New York: Oxford University)

\bibitem{Pethick_7}
Pethick C J and Smith H 2008 
{\it Bose-Einstein Condensation in Dilute Gases} 
(Cambridge: Cambridge University)

\bibitem{Courteille_8}
Courteille P W, Bagnato V S and Yukalov V I 2001 
{\it Laser Phys.} {\bf 11} 659

\bibitem{Andersen_9}
Andersen J O 2004 
{\it Rev. Mod. Phys.} {\bf 76} 599

\bibitem{Yukalov_10}
Yukalov V I 2004 
{\it Laser Phys. Lett.} {\bf 1} 435

\bibitem{Bongs_11}
Bongs K and Sengstock K 2004 
{\it Rep. Prog. Phys.} {\bf 67} 907

\bibitem{Yukalov_12}
Yukalov V I and Girardeau M D 2005 
{\it Laser Phys. Lett.} {\bf 2} 375

\bibitem{Posazhennikova_13}
Posazhennikova A 2006 
{\it Rev. Mod. Phys.} {\bf 78} 1111

\bibitem{Yukalov_14}
Yukalov V I 2007 
{\it Laser Phys. Lett.} {\bf 4} 632

\bibitem{Proukakis_15}
Proukakis N P and Jackson B 2008 
{\it J. Phys. B} {\bf 41} 203002

\bibitem{Yurovsky_16}
Yurovsky V A, Olshanii M and Weiss D S 2008 
{\it Adv. At. Mol. Opt. Phys.} {\bf 55} 61

\bibitem{Yukalov_17}
Yukalov V I 2009 
{\it Laser Phys.} {\bf 19} 1

\bibitem{Yukalov_18}
Yukalov V I 2011 
{\it Phys. Part. Nucl.} {\bf 42} 460

\bibitem{Yukalov_19}
Yukalov V I, Yukalova E P and Bagnato V S 1997 
{\it Phys. Rev. A} {\bf 56} 4845

\bibitem{Yukalov_47}
Yukalov V I, Yukalova E P and Bagnato V S 2000 
{\it Laser Phys.} {\bf 10} 26 

\bibitem{Seman_20}
Seman J A, Henn E A L, Haque M, Shiozaki R F, Ramos E R F, Caracanhas M, 
Castilho P, Castelo Branco C, Tavares P E S, Poveda-Cuevas F J, Roati G, 
Magalh\~{a}es K M F and Bagnato V S 2010 {\it Phys. Rev. A} {\bf 82} 033616 

\bibitem{Yukalov_21}
Yukalov V I, Yukalova E P and Bagnato V S 2002 
{\it Phys. Rev. A} {\bf 66} 043602

\bibitem{Yukalov_22}
Yukalov V I 1980 
{\it Acta Phys. Pol. A} {\bf 57} 295

\bibitem{Dutton_23}
Dutton Z, Budde M, Slowe C and Hau L V 2001 
{\it Science} {\bf 293} 663

\bibitem{Yukalov_24}
Yukalov V I and Yukalova E P 2004 
{\it Laser Phys. Lett.} {\bf 1} 50

\bibitem{Ruostekoski_25}
Ruostekoski J and Dutton Z 2005 
{\it Phys. Rev. A} {\bf 72} 063626

\bibitem{Shomroni_26}
Shomroni I, Lahoud E, Levy S and Steinhauer J 2009 
{\it Nature Phys.} {\bf 5} 193

\bibitem{Ma_27}
Ma M, Carretero-Gonzalez R, Kevrekidis P G, Frantzeskakis D J and Malomed B A 2010 
{\it Phys. Rev. A} {\bf 82} 023621

\bibitem{Ishiro_28}
Ishiro S, Tsubota M and Takeuchi H 2011 
{\it Phys. Rev. A} {\bf 83} 063602

\bibitem{Simula_29}
Simula T P 2011 
{\it Phys. Rev. A} {\bf 84} 021603

\bibitem{Henn_30}
Henn E A L, Seman J A, Roati G, Magalh\~{a}es K M F and Bagnato V S 2009 
{\it Phys. Rev. Lett.} {\bf 103} 045301

\bibitem{Seman_31}
Seman J A, Shiozaki R F, Poveda-Cuevas F J, Henn E A L, Magalh\~{a}es K M F, 
Roati G, Telles G D and Bagnato V S 2011 
{\it J. Phys. Conf. Ser.} {\bf 264} 012004

\bibitem{Shiozaki_32}
Shiozaki R F, Telles G D, Yukalov V I and Bagnato V S 2011 
{\it Laser Phys. Lett.} {\bf 8} 393

\bibitem{Seman_33}
Seman J A, Henn E A L, Shiozaki R F, Roati G, Poveda-Cuevas F J, Magalh\~{a}es K M F, 
Yukalov V I, Tsubota M, Kobayashi M, Kasamatsu K and Bagnato V S 2011 
{\it Laser Phys. Lett.} {\bf 8} 691

\bibitem{Bagnato_48}
Bagnato V S and Yukalov V I 2013
{\it Prog. Opt. Sci. Photon.} {\bf 1} 377

\bibitem{Henn_34}
Henn E A L, Seman J A, Ramos E R F, Caracanhas M, Castilho P, Olimpio E P, Roati G, 
Magalh\~{a}es D V, Magalh\~{a}es K M F and Bagnato V S 2009 
{\it Phys. Rev. A} {\bf 79} 043618

\bibitem{Ramos_35}
Ramos E R F, Henn E A L, Seman J A, Caracanhas M A, Magalh\~{a}es K M F, Helmerson K, 
Yukalov V I and Bagnato V S 2008 
{\it Phys. Rev. A} {\bf 78} 063412

\bibitem{Pollack_36}
Pollack S E, Dries D, Hulet R G, Magalh\~{a}es K M F, Henn E A L, Ramos E R F, 
Caracanhas M A and Bagnato V S 2010 
{\it Phys. Rev. A} {\bf 81} 053627 

\bibitem{Yukalov_40}
Yukalov V I 1991 
{\it Phys. Rep.} {\bf 208} 395

\bibitem{Yukalov_41}
Yukalov V I 2003 
{\it Int. J. Mod. Phys.} {\bf 17} 2333 

\bibitem{Zakharov_42}
Zakharov V E, Lvov V S and Falkovich G E 1992 
{\it Kolmogorov Spectra of Turbulence - Wave Turbulence} 
(Berlin: Springer)

\bibitem{Kibble_43}
Kibble T W 1976 
{\it J. Phys. A} {\bf 9} 1387 

\bibitem{Zurek_44}
Zurek W H 1996 
{\it Phys. Rep.} {\bf 276} 177 

\bibitem{Yukalov_45}
Yukalov V I, Yukalova E P and Bagnato V S 2009 
{\it Laser Phys.} {\bf 19} 686

\bibitem{Yukalov_46}
Yukalov V I and Graham R 2007 
{\it Phys. Rev. A} {\bf 75} 023619

\end{thebibliography}
\end{document}